\documentclass[%
 reprint,
superscriptaddress,
amsmath,amssymb,
aps,
prb,
]{revtex4-2}
\usepackage[english]{babel}
\usepackage{subfigure}
\usepackage{graphicx}
\usepackage{dcolumn}
\usepackage{bm}
\usepackage{hyperref}

\usepackage{svg}
\usepackage{hyperref}
\usepackage{cleveref}
\usepackage{amssymb}
\usepackage{pifont}

\begin{document}

\title{3/8 charge-density wave instability in the kagome metal CsV$_3$Sb$_5$}

\author{Luca Buiarelli}
\thanks{These authors contributed equally to this work.}
\affiliation{Department of Chemical Engineering and Materials Science, University of Minnesota, MN 55455, USA}

\author{Hyeonseo H. Park}
\thanks{These authors contributed equally to this work.}
\affiliation{Department of Chemical Engineering and Materials Science, University of Minnesota, MN 55455, USA}

\author{Ethan T. Ritz}
\affiliation{Department of Engineering, Harvey Mudd College, Claremont, CA 91711, USA}
\affiliation{Department of Civil and Mineral Engineering, University of Toronto, Toronto, ON M5S 1A7, Canada}

\author{Rafael M. Fernandes}
\affiliation{Department of Physics, The Grainger College of Engineering, University of Illinois Urbana-Champaign, Urbana, IL 61801, USA}
\affiliation{Anthony J. Leggett Institute for Condensed Matter Theory, The Grainger College of Engineering, University of Illinois Urbana-Champaign, Urbana, IL 61801, USA}

\author{Turan Birol}
\thanks{Corresponding author: \href{mailto:tbirol@umn.edu}{tbirol@umn.edu}}
\affiliation{Department of Chemical Engineering and Materials Science, University of Minnesota, MN 55455, USA}

\date{\today}

\begin{abstract}
The kagome metal CsV$_3$Sb$_5$ exhibits charge density wave (CDW) and superconducting transitions, both of which are substantially affected by pressure. Recent x-ray diffraction experiments identify a new charge-ordered phase at a pressure coinciding with the dip of the superconducting dome. This CDW phase displays a distinctive wavevector $\mathbf{Q} = (3/8, 0, 1/2)$ and monoclinic symmetry, in contrast to the $2\times2\times 2$ and $2\times2\times 4$ orders reported at ambient pressure. In this letter, we show that density functional perturbation theory (DFPT) calculations on a fine reciprocal-space grid predict the leading lattice instability of CsV$_3$Sb$_5$ to be at this wavevector. Finite-displacement calculations reveal that while anharmonic effects stabilize the conventional $L_2^-$ CDW at ambient pressure, the competing $3/8$ instability becomes energetically favorable above $\sim1$~GPa, consistent with experimental observations. Interestingly, the nesting function displays a peak at the same wavevector. A Landau free energy analysis explains the emergence of the monoclinic distortion observed in x-ray diffraction, in addition to several bond-ordering patterns that may be relevant to the observed trends in the superconducting $T_c$. 
\end{abstract}

\maketitle

\textit{Introduction.---}
The kagome metals AV$_3$Sb$_5$ (A = K, Rb, Cs) are well-established hosts of a rich phase diagram, encompassing superconductivity (SC), charge-density wave (CDW) order, and possible loop-current phases whose precise nature remains under active investigation~\cite{Jiang2022Sep, Neupert2022Feb, Wilson2024Jun, DiSante2026, Fernandes2026}. These compounds share the common features of a CDW transition at $T_{\rm CDW} \approx 80$--$100$~K, and a SC transition at $T_c \approx 1$--$3$~K, with both transition temperatures impacted by the alkali metal A~\cite{Ortiz2019, Ortiz2020Dec,Ortiz2021Mar, Yin2021Mar}. The CDW state and the accompanying change in the crystal structure are primarily associated with a bond order between the V ions that form the kagome lattice, and at least double the unit cell in all three directions, lowering the symmetry from hexagonal to orthorhombic~\cite{Kautzsch2023}. This structural distortion is linked to unstable phonon modes transforming as the $M_1^+$ and $L_2^-$ irreducible representations (irreps) at the zone-boundary points $M=(1/2,1/2,0)$ and $L=(1/2,1/2,1/2)$~\cite{Ratcliff2021Nov,Tan2021Jul,Park2021,Christensen2021Dec}. In the temperature range between the onset of CDW order and the superconducting transition, a range of unusual phenomena have been reported, particularly under the application of strain or magnetic field \cite{Guo2024}. For instance, there have been reports of electronic nematicity developing within the CDW state, with STM and ARPES measurements revealing unidirectional charge modulations and threefold symmetry breaking across the family of kagome metals~\cite{Zhao2021Sep, Xu2022Three-state,Li2022, Jiang2023, Wu2023}. Moreover, phenomena consistent with time-reversal symmetry-breaking have also been observed in muon spin-rotation~\cite{Mielke2022, Guguchia2023,Bonfa2025}, optics \cite{Xu2022Three-state}, magnetic torque~\cite{Asaba2024,Gui2025}, STM~\cite{Jiang2021,Xing2024}, and transport \cite{Guo2022}. While these results are suggestive of broken-symmetry phases, their interpretation remains under debate \cite{HHWen2022,Saykin2023Kerr,Farhang2023,JHChu2024}. Theoretically, several different mechanisms have been proposed that yield spontaneous threefold and time-reversal symmetry breaking \cite{Park2021,Denner2021,Lin2021,JPHu2021,Christensen2022,Tazai2022,Fischer2023,Ziqiang2023,HYKee2024}. 

\begin{figure*}[]
    \centering
    \includegraphics[width=0.95\linewidth]{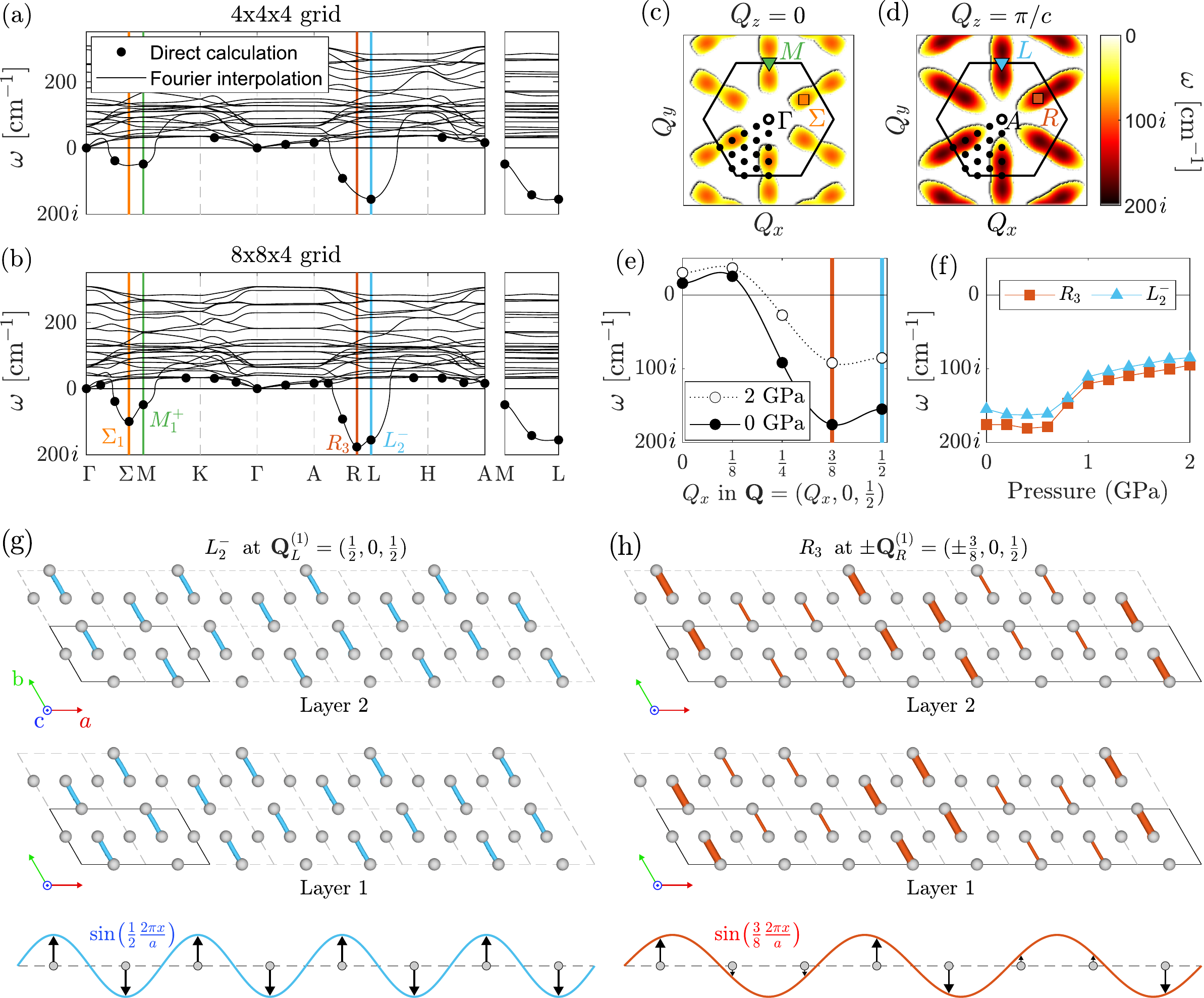}
    \caption{Phonon instabilities of CsV$_3$Sb$_5$ from DFPT. (a--b) Phonon dispersion along high-symmetry lines, calculated on a $4\times4\times4$ (a) and $8\times8\times4$ (b) $\mathbf{Q}$-point grid; black circles show frequencies computed directly by DFPT. The finer grid in (b) reveals the $\Sigma_1$ and $R_3$ instabilities absent in (a). (c--d) Color plot of the imaginary part of the unstable phonon frequencies in the Brillouin zone of $P6/mmm$ at the high-symmetry planes (c) $Q_z = 0$ and (d) $Q_z = \pi/c$; yellow and orange squares mark the points along the $\Sigma$ and $R$  lines with $\delta = 3/8$. (e) Unstable phonon frequencies along the $A$--$L$ line at applied pressures of $0$ and $2$~GPa. (f) Pressure evolution of the $R_3$ and $L_2^-$ imaginary frequencies, showing slow hardening above $\sim0.8$~GPa. (g--h) Real-space displacement patterns of the $L_2^-$ mode at $\mathbf{Q}_L^{(1)} = (1/2, 0, 1/2)$ and the $R_3$ mode at $\pm\mathbf{Q}_R^{(1)} = (\pm3/8, 0, 1/2)$, shown on two consecutive kagome layers. Both modes alternate sign along $c$, but while $L_2^-$ doubles the unit cell along $a$, $R_3$ requires an eightfold expansion.}
    \label{fig:phonons}
\end{figure*}

CsV$_3$Sb$_5$ stands out among the kagome metals for its particularly complex phase diagram. In addition to the dominant $2\times2\times2$ CDW, it has been shown to host fragile coexisting CDW distortions that enlarge the unit cell to $2\times2\times4$~\cite{Stahl2022,Xiao2023Mar, Plumb2024Sep}. Moreover, its SC transition temperature displays a double-dome structure as a function of both pressure and hole doping~\cite{Chen2021Jun, Yu2021Jun, Oey2022Apr}. Interestingly, a recent high-resolution x-ray diffraction study found that along the pressure range $0.7\,\mathrm{GPa}<P<1.7\,\mathrm{GPa}$, where the SC double-dome is suppressed (but with $T_c$ still nonzero), a Bragg peak at wavevector $\mathbf{Q} = (3/8, 0, 1/2)$ emerges, with the associated structure determined to be monoclinic (space group \#12, $C2/m$)~\cite{Stier2024Dec}. Previous nuclear magnetic resonance~\cite{Zheng2022emergent} and nuclear quadrupole resonance~\cite{Feng2023commensurate} experiments identified new spectral features in the same pressure range, attributing them to a stripe-like unit-cell quadrupling or incommensurate modulations. Similarly, short-range unidirectional CDW correlations near $\mathbf{Q} \approx (3/8, 0, 1/2)$ were also observed in hole-doped CsV$_3$Sb$_{5-x}$Sn$_x$~\cite{Kautzsch2023Jul, Salinas2025}, again in a region coinciding with a suppression of the the superconducting $T_c$.

In this Letter, by using first principles density functional theory and Landau free-energy phenomenology, we show that the strongest CDW instability of the high temperature phase of CsV$_3$Sb$_5$ has wavevectors away from the high-symmetry zone boundary points $M$ and $L$. In particular, by employing a sufficiently dense momentum grid, we find the strongest harmonic instability to be the $R_3$ mode with wavevector near $\mathbf{Q_R}=(3/8,0,1/2)$, consistent with the experimental observations discussed above. While within the harmonic approximation the $R_3$ mode is the most unstable, we find that anharmonic effects make the previously studied $L_2^-$ mode energetically favorable at ambient pressure. Conversely, the $R_3$ mode becomes the energetically favored instability for pressures above $\approx 1$~GPa, consistent with the experimental results. Finally, we show that the Fermi surface nesting function has a local peak close to the same $\mathbf{Q_R}$ wavevector, and eloborate on the reconstruction of the electronic structure by the $R_3$ distortion.

\textit{Phonon Instabilities.---} In order to map out the lattice instabilities in the high symmetry high temperature phase of CsV$_3$Sb$_5$, we start by performing density functional perturbation theory (DFPT) calculations~\cite{Gonze1997Apr} (see the  Supplemental Material~\cite{Supplement} for details). In a typical first principles calculation of phonon dispersions, a set of single $\mathbf{Q}$-point DFPT calculations are performed to obtain the dynamical matrix $\mathcal{D}(\mathbf{Q})$ on a grid of wavevectors. This information is then used to obtain real-space force constants, which are Fourier-transformed back to obtain $\mathcal{D}(\mathbf{Q})$ at arbitrary $\mathbf{Q}$ points~\cite{Baroni2001} \footnote{If the frozen phonons approach is used, then the real space force constants can be directly obtained from a calculation in a supercell.}. Therefore, in the phonon dispersion plots,  the frequencies for arbitrary wavevectors located in-between the grid points are reliable only as long as $\mathcal{D}(\mathbf{Q})$ behaves smoothly, or equivalently, long-range interatomic force-constants in real space are negligable \cite{Gonze1997Apr}. 
Previous calculations on CsV$_3$Sb$_5$ have employed $\mathbf{Q}$-grids (or equivalent supercells) of $3\times3$~\cite{Tan2021Jul, Ratcliff2021Nov, He2024, Bhandari2024, Chen2025} or $6\times6$~\cite{Zhang2021} in-plane, all of which consistently identified the $L_2^-$ mode as the leading instability, i.e., the imaginary-frequency mode with the largest magnitude. However, the wavevector $\mathbf{Q}_R^{(1)} = (3/8, 0, 1/2)$ is commensurate only with grids whose in-plane dimension is a multiple of 8, meaning that it is absent from these previously used grids. 

Our DFPT results shown in Figs.~\ref{fig:phonons}(a--b), which compare phonon dispersions obtained from $4\times4\times4$ and $8\times8\times4$ $\mathbf{Q}$-grids, underline the importance of the grid density. Note that, as it is customary when plotting unstable phonon modes, frequencies located on the ``negative'' vertical axis are actually imaginary; the most unstable mode is then assigned to the frequency whose imaginary part has the largest magnitude.  While both grids yield unstable modes at the zone-boundary $M=(1/2, 0, 0)$ and $L=(1/2, 0, 1/2)$ points, transforming as the $M_1^+$ and $L_2^-$ irreducible representations of $P6/mmm$ (\#191), the most unstable frequency in the finer grid (Fig.~\ref{fig:phonons}(b)) is not at the $L$ point (like in Fig.~\ref{fig:phonons}(a)), but on lower-symmetry $\mathbf{Q}$-points along $\Gamma$--$M$ ($\Sigma$ line) and $A$--$L$ ($R$ line). Here, $\Gamma=(0,0,0)$ and $A=(0,0,1/2)$. 

This behavior is also evident in Figs.~\ref{fig:phonons}(c--d), where we show in a color plot the amplitude of the unstable phonon frequencies on the $Q_z=0$ and $Q_z=\pi/c$ planes of the Brillouin zone. On both planes, the most unstable frequency, which is indicative of the wavevector of the leading instability, is not on the high-symmetry points, but along the $\Sigma$ and $R$ lines. 
The unstable modes transform as the $\Sigma_1$ and $R_3$ irreps, and are located at momenta $\mathbf{Q}_\Sigma^{(1)} = (\delta, 0, 0)$ and $\mathbf{Q}_R^{(1)} = (\delta, 0, 1/2)$ and the symmetry-equivalent wavevectors in their stars, with $\delta \simeq 3/8=0.375$, (see the End Matter). The $R_3$ mode is the leading (most unstable) instability in the entire Brillouin zone, meaning that it has the largest imaginary part. 

We note that if we use the full Fourier-interpolated dispersion, the wavevector of the most unstable mode moves to $\delta \approx 0.39$. However, it is not possible to confirm this using DFPT due to the limitations of the $\mathbf{Q}$-grids that may be practically employed (as discussed above). Similarly, performing a frozen phonon calculation at this wavevector would require prohibitively large supercells.
Conversely, since $\mathbf{Q}_R^{(1)}$ lies directly on the $8\times8\times4$ grid, the frequency at this point is an outcome of a single DFPT calculation at this wavevector, rather than a possible artifact of Fourier interpolation. In Figs.~\ref{fig:phonons}(a--b), we display the frequencies computed directly by DFPT as black circles. 
In the remainder of this study, we focus on $\mathbf{Q}_R^{(1)} = (\delta=3/8, 0, 1/2)$, and simply refer to the unstable mode at this wavevector as \textit{the} $R_3$ mode. 

Fig.~\ref{fig:phonons}(g--h) contrasts the real-space characters of the $L_2^-$ and $R_3$  modes extracted from their polarizations. Apart from the different periodicities (2 unit cells along [100] for $L_2^-$, as opposed to 8 unit cells for $R_3$), both modes are well described as nearest-neighbor bond order between the V ions. 

The pressure dependence of the unstable phonon branch near $L$ is shown in Figs.~\ref{fig:phonons}(e--f). Both the $R_3$ and $L_2^-$ instabilities exhibit only a small change in frequency up to $\sim 0.8$~GPa, above which both of their frequencies decrease in absolute magnitude sharply, and continue to decrease with pressure afterwards. Nevertheless, both modes remain unstable well beyond $2$~GPa, consistent with earlier calculations~\cite{Ritz2023May}. This relatively weak sensitivity to pressure at the harmonic level suggests that the  emergence of a $R_3$ instability in pressure experiments must be driven by anharmonic effects, which we study next.

\begin{figure}[]
    \centering
    \includegraphics[width=0.99\linewidth]{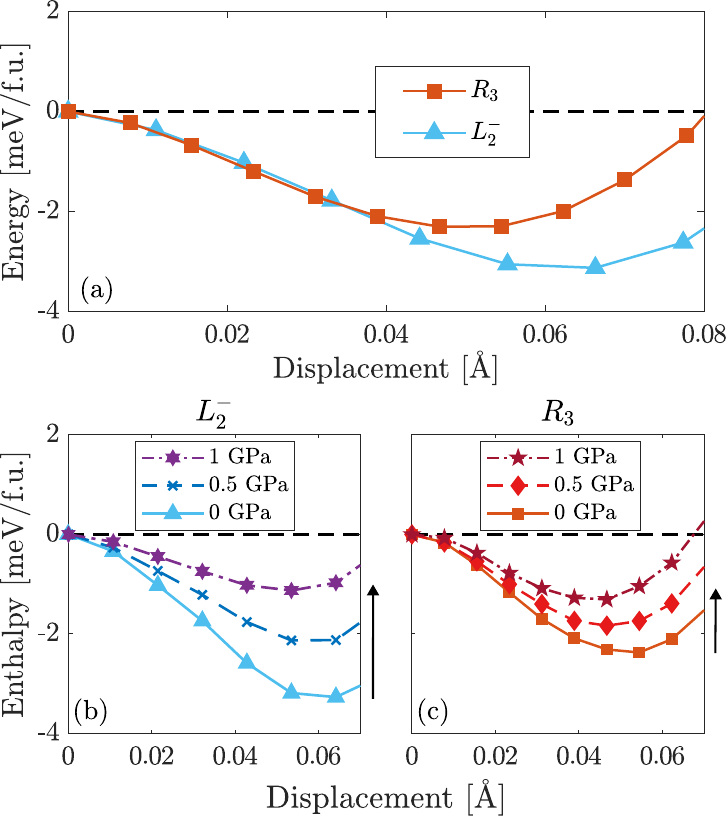}
    \caption{(a) Total energy as a function of the displacement amplitude along the $L_2^-$ and $R_3$ phonon eigenvectors, computed in $2\times1\times2$ and $8\times1\times2$ supercells, respectively. 
    The displacement on the horizontal axis is the root‑sum‑of‑squares displacement normalized by the square root of the supercell size relative to the primitive cell.
    (b--c) Enthalpy along the same distortions at applied pressure values of $0$, $0.5$, and $1$~GPa. The contrasting pressure sensitivity of the two modes is the key result: the $L_2^-$ minimum rises rapidly with pressure (b) while the $R_3$ minimum remains comparatively stable (c), switching the most stable phase near $1$~GPa.
    \label{fig:supercell}}
\end{figure}

\textit{Anharmonic effects.---} The harmonic approximation alone cannot determine which instability condenses at low temperatures, as it does not capture the higher-order terms  that may ultimately determine the ground state. Earlier first principles work~\cite{Ritz2023May} showed that trilinear coupling terms between the $M_1^+$ and $L_2^-$ modes can lower the energy by $\sim10$~meV/f.u. compared to that of a single-$\mathbf{Q}$ CDW phase, favoring the observed staggered tri-hexagonal phase. As we discuss below, analogous trilinear couplings exist between the $\Sigma_1$ and $R_3$ order parameters. Unfortunately, studying the couplings between these modes requires a prohibitively large supercell. Instead, we focus on the single-mode anharmonic effects by performing finite-displacement DFT calculations in $8\times1\times2$ supercells, displacing ions by either of the two unstable modes independently.

The Kohn-Sham energies as a function of displacement are shown in Fig.~\ref{fig:supercell}(a). Both $L_2^-$ and $R_3$ exhibit negative curvatures at the origin, consistent with the imaginary frequencies of $179$~cm$^{-1}$ and $200$~cm$^{-1}$ for $L_2^-$ and $R_3$ shown in Fig.~\ref{fig:phonons}(b). Despite the larger imaginary frequency of $R_3$ suggesting it to be the leading instability at the harmonic level, the higher-order (anharmonic) terms lead to a deeper energy minimum for $L_2^-$.  This is in agreement with experiments, which find an $L_2^-$ instability at ambient pressure.

To assess whether the ground state changes to $R_3$  under pressure, we plot
in Figs.~\ref{fig:supercell}(b--c) the enthalpy as a function of either distortion pattern at different pressures. Because these calculations with fixed-pressure boundary conditions involve relaxing the lattice parameters, enthalpy is a more relevant quantity rather than energy~\cite{Supplement}. The enthalpy minimum of the  $L_2^-$  mode rises rapidly with pressure ($\sim 2$~meV/(GPa$\cdot$f.u.)), whereas the $R_3$ enthalphy minimum is far less sensitive ($\sim 1$~meV/(GPa$\cdot$f.u.)). As a result, by $1$~GPa, the $R_3$ enthalpy minimum already lies below that of $L_2^-$, consistent with the single-$\mathbf{Q}$ $R_3$ phase emerging only above approximately $0.7$~GPa in the experiment~\cite{Stier2024Dec}. 
%

\textit{Landau theory.---} 
To understand the symmetry-allowed couplings between the different possible CDW orders, we now construct the Landau free energy of the order parameters corresponding to all four leading unstable phonon modes: $\mathbf{M} = (m_1, m_2, m_3)$, $\mathbf{L} = (l_1, l_2, l_3)$, $\boldsymbol{\Sigma} = (\sigma_1, \sigma_2, \sigma_3)$, and $\mathbf{R} = (r_1, r_2, r_3)$. Each order parameter transforms as the irreducible representation of the unstable phonon mode at the corresponding wavevector $\mathbf{Q}_M^{(i)}$, $\mathbf{Q}_L^{(i)}$, $\mathbf{Q}_\Sigma^{(i)}$, $\mathbf{Q}_R^{(i)}$, defined in the Supplemental Material. Since $M$ and $L$ lie at time-reversal invariant momenta (TRIM) points, where $-\mathbf{Q}$ is equivalent to $\mathbf{Q}$ modulo a reciprocal lattice vector, $m_i$ and $l_i$ are real-valued. In contrast, $\Sigma$ and $R$ do not lie at TRIM points, since $-\mathbf{Q}_\Sigma^{(i)}$ and $-\mathbf{Q}_R^{(i)}$ are not equivalent to $\mathbf{Q}_\Sigma^{(i)}$ and $\mathbf{Q}_R^{(i)}$. Consequently, $\sigma_i$ and $r_i$ are complex, with $\sigma_i^*$ and $r_i^*$ being the independent order parameters at the time-reversal partner wavevectors. Up to third order, there is no mixing between terms that include $\mathbf{M}$ and $\mathbf{L}$, and those that include $\mathbf{R}$ and $\mathbf{\Sigma}$. Hence, the free energy can be written as $\mathcal{F} = \mathcal{F}_{ML} + \mathcal{F}_{R\Sigma}$. 

The form of $\mathcal{F}_{ML}$ has been studied before~\cite{Park2021, Christensen2021Dec, Ritz2023May}. The most unusual part of it is the trilinear couplings $m_1 m_2 m_3$ and $m_1 l_2 l_3$ (plus permutations), which give rise to phases such as the staggered tri-hexagonal phase $(MLL)$, where two components of the $\mathbf{L} $ order parameter  mix with one component of the $\mathbf{M} $ order parameter~\cite{Christensen2021Dec}. Other relevant phases in the phase diagram of $\mathcal{F}_{ML}$ include the single-$\mathbf{Q}$ $(M00)$ and $(L00)$ phases, as well as the triple-$\mathbf{Q}$ $(MMM)$ and $(LLL)$. 

Similar to $\mathcal{F}_{ML}$, $\mathcal{F}_{R\Sigma}$ also has two trilinear coupling terms involving $\mathbf{\Sigma}$ and $\mathbf{R}$
\begin{equation}
\begin{split}
\label{eq:free_energy_rs}
    \mathcal{F}_{R\Sigma} =\; &\frac{\alpha_\Sigma}{2} |\boldsymbol{\Sigma}|^2 + \frac{\alpha_R}{2} |\mathbf{R}|^2+ \frac{\gamma_\Sigma}{3} \,\mathbf{Re}(\sigma_1 \sigma_2^* \sigma_3) \\
    &+ \frac{\gamma_{\Sigma R}}{3} \,\mathbf{Re}(\sigma_1 r_2^* r_3 + r_1 \sigma_2^* r_3 + r_1 r_2^* \sigma_3)\,,
\end{split}
\end{equation}
where $|\boldsymbol{\Sigma}|^2 = |\sigma_1|^2 + |\sigma_2|^2 + |\sigma_3|^2$ and  $|\mathbf{R}|^2=|r_1|^2 + |r_2|^2 + |r_3|^2$. 
The trilinear coupling $\gamma_{\Sigma R}$ makes the simultaneous condensation of $\Sigma_1$ and $R_3$ energetically favorable and it is the analogue of the $M_1^+$--$L_2^-$ coupling that stabilizes the staggered tri-hexagonal phase $(MLL)$ at ambient pressure \cite{Ritz2023May}. The cross-coupling between the two sets of order parameters ($M_1^+$--$L_2^-$ and $\Sigma_1$--$R_3$) enters only at fourth order through quadrilinear and biquadratic terms detailed in the SM~\cite{Supplement}, which justifies focusing on the possible phases that emerge from $\mathcal{F}_{R\Sigma}$ only. 

However, despite the similarities in the free energy expressions  $\mathcal{F}_{ML}$ and $\mathcal{F}_{R\Sigma}$, the ground state manifold of the latter is larger because of the complex nature of these order parameters. For example, in analogy with the single-$\mathbf{Q}$ $(L00)$ phase, one can consider the $(R00)$ phase. Depending on the phase of the complex order parameter, i.e. the relative amplitude of the distortions with time-reversed wavevectors $(\pm 3\pi/8, 0, 1/2)$, the space group can be lowered to either $Cmmm$ (\#65, for real $r_1$) or $Amm2$ (\#38, for complex $r_1$). The energies of these phases are equal in the Landau free energy up to 6$^{\textrm{th}}$ order if couplings with other modes (including strain) are not taken into account. Similarly, the triple-$\mathbf{Q}$ $(RRR)$ phase can lower the space group to either $P6/mmm$ (\#191) or $P6_3/mmc$ (\#194), depending on the relative phases of the components. We provide a complete list of possible low-symmetry phases that can be constructed from the four order parameters in the Supplementary Tables S1-4~\cite{Supplement}. 
%

To further constrain the combinations realized experimentally, we use the fact that the experimentally reported space group is the monoclinic $C2/m$~\cite{Stier2024Dec}. This symmetry can be realized through 
two distinct combinations of order parameters: The first possibility is the condensation of two unequal $R_3$ components, which necessarily induces a $\Sigma_1$ component due to the trilinear term in Eq.~\ref{eq:free_energy_rs}. We refer to this phase as $(\Sigma RR^\prime)$. The second possibility is the condensation of a single $R_3$ component within the ambient-pressure $(MLL)$ phase, $(MLL)+(R00)$. 
%
Without performing additional calculations to obtain all the coefficients in $\mathcal{F}$, it is not possible to predict which one of the two combinations of order parameters is favorable. This is not practical due to the large ($8\times 8 \times 2$) supercells required. 
%
%
However, the stability of the ($MLL$) at ambient pressure and the observation of a coexistence region between the different CDW phases \cite{Stier2024Dec} suggest that ($MLL$)+($R00$) phase is more likely.
This phase can be understood by starting from deep inside the ($MLL$) phase, at the mean-field level, and taking into account the fourth order terms in the free energy to obtain an effective free energy for the $R_3$ mode:
\begin{equation}
\begin{split}
    \mathcal{F}_R^{\rm \,eff} \sim\; &\frac{\alpha_{R*}}{2}|\mathbf{R}|^2 
    + \frac{\alpha'_{R*}}{2}|r_1|^2 + \frac{u_{R}}{4}|\mathbf{R}|^4.
\end{split}
\end{equation}
The effect of the $M$ and $L$  distortions is not only to renormalize the bare $\alpha_R$ through biquadratic couplings, but also to lift the three-fold degeneracy between $r_1$, $r_2$, $r_3$ through the $\alpha'_{R*}$ term. If $\alpha'_{R*}<0$, this picks a single component of $R$ to condense if the higher order terms don't compete against it, yielding the monoclinic $C2/m$ symmetry.

\textit{Electronic structure.---} 
\begin{figure}[]
    \centering
    \includegraphics[width=0.95\linewidth]{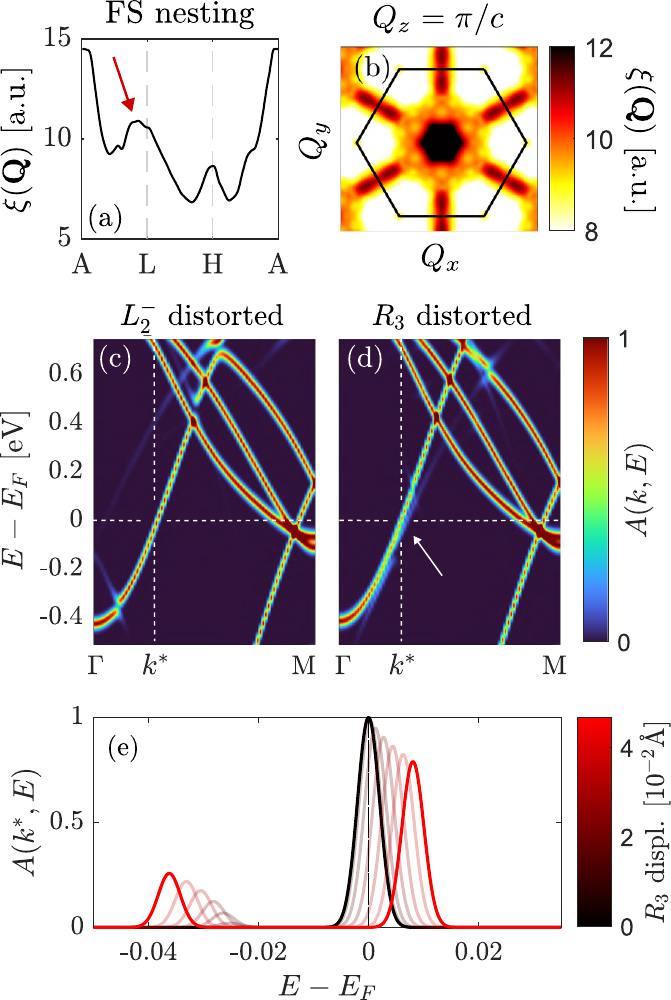}
    \caption{Electronic structure of CsV$_3$Sb$_5$ and its response to the $R_3$ distortion. (a) Fermi surface nesting function $\xi(\mathbf{Q})$ along the $A$--$L$--$H$--$A$ path and (b) color plot of $\xi(\mathbf{Q})$ in the Brillouin zone at $Q_z = \pi/c$. The latter reveals a momentum-dependent structure reminiscent of that of the unstable phonon modes shown in Fig. \ref{fig:phonons}(d). (c--d) Band structure (spectral density) unfolded into the primitive Brillouin zone along the $\Gamma$--$M$ line for the $L_2^-$ distorted structure at its energy minimum (c) and the $R_3$ distorted structure at its energy minimum (d). (e) Energy cuts of the spectral density as a function of $R_3$ displacement, at fixed momentum $k^*$. }
    \label{fig:unfolded}
\end{figure}
We now investigate which, if any, electronic structure features are related with the stronger instability at the $R$ and $\Sigma$ points. While a direct connection between Fermi surface nesting and the CDW instability wavevector cannot be made~\cite{Johannes2008}, the nesting function $\xi(\mathbf{Q})$ (Eq.~\ref{eq:nesting} in End Matter) can nevertheless provide useful insights.
In order to calculate the nesting function in CsV$_3$Sb$_5$ with high accuracy, we used Wannier interpolation of the DFT electronic structure (see End Matter) and a $1000\times 1000$ $\mathbf{Q}$-point grid for each value of $Q_z$. Our results agree with results available in the literature~\cite{Kaboudvand2022, Wu2022Apr}. However, we highlight a feature not explicitly emphasized before: on the $Q_z=\pi/c$ plane, the susceptibility has a local maximum near the $3/8$ wavevector, as shown in Fig.~\ref{fig:unfolded}(a). Moreover, near the Brillouin zone edges, the momentum-dependence of the susceptibility, shown in Fig.~\ref{fig:unfolded}(b) with peaks along the $A$-$L$ direction, closely resembles that of the unstable phonon modes in Fig.~\ref{fig:phonons}(d). Interestingly, this is not the case on the $Q_z=0$ plane, where the nesting peaks are along the $M$–$K$ line (Fig.~\ref{fig:nesting}).

Experimentally, the superconducting  $T_c$ deep inside the $3/8$ phase is observed to be substantially suppressed with respect to the $T_c$ inside the period-2 phase. To gain further insight into this issue, we analyze the effect of the $R_3$ crystalline distortion on the electronic structure by calculating the DFT bandstructure in the $R$ phase, and then unfolding the bands along the $\Gamma$--$M$ line from the reduced Brillouin zone of the $8\times1\times2$ distorted supercell back to the Brillouin zone of the primitive cell. The results are shown in Fig.~\ref{fig:unfolded}(c--d), where we compare the undistorted electronic structure with the reconstructed bands under the $L_2^-$ and $R_3$ distortions. The strongest spectral weight redistributions in both the $L_2^-$ and $R_3$ phases occur in the vanadium $d$-bands approximately $0.6$~eV above the Fermi level. However, one of the bands folded by the $R_3$ order parameter crosses the Fermi level near the existing antimony-derived $\Gamma$ Fermi pocket, reducing its spectral weight. Energy cuts of the spectral density at fixed momentum $k$ as a function of the $R_3$ distortion, shown in Fig.~\ref{fig:unfolded}(e), makes this effect more apparent. As the atoms are displaced, spectral weight is transferred from the original peak to a new peak appearing below the Fermi level. This gives rise to the smeared-like band near the Fermi level shown in Fig.~\ref{fig:unfolded}(d). Previous experimental and theoretical work have identified the Sb-derived $\Gamma$ pocket as an important ingredient for superconductivity in CsV$_3$Sb$_5$~\cite{Tsirlin2022, Tsirlin2023, Ritz2023Sep, Schultz2026}. While further theoretical investigations are needed, it is plausible that the redistribution of spectral weight near the Fermi level inside the 3/8 phase could contribute to the observed suppression of the superconducting $T_c$.

\textit{Conclusions.---}
In this work, we showed that first-principles calculations with sufficient reciprocal-space resolution reveal that the leading harmonic instability in CsV$_3$Sb$_5$ is the $R_3$ mode with wavevector $\mathbf{Q} = (3/8, 0, 1/2)$.  Our    finite-displacement supercell calculations demonstrate that anharmonic terms favor the conventional $L_2^-$ CDW at ambient pressure. As pressure is applied, the enthalpy of the $L_2^-$  phase increases faster than that of the $R_3$ phase, driving a change in their relative stability near $1$~GPa that is consistent with the experimentally observed phase boundary. Analysis of the unfolded electronic structure shows that the $R_3$ distortion causes the Sb-derived $\Gamma$ pocket to lose spectral weight due to band folding, which might be connected to the suppression of $T_c$ observed experimentally in the intermediate pressure range. A Landau free energy analysis reveals trilinear couplings between $R_3$ and its $Q_z=0$ counterpart $\Sigma_1$, suggesting that multi-$\mathbf{Q}$ phases at $3/8$ momentum are symmetry-allowed, although none have been reported experimentally to date.

Several open questions remain. Whether the true instability wavevector is commensurate at $\delta = 3/8$ or slightly incommensurate cannot be resolved within the supercell sizes accessible to current calculations. The nature of the fully coupled $3/8$ ground state, including the possible role of an induced secondary order parameter, similarly requires mapping the potential energy surface beyond the single-mode approximation studied here. Finally, the apparent similarity between the pressure-driven and hole-doping-driven $3/8$ phases in CsV$_3$Sb$_{5-x}$Sn$_x$ suggests that a unified understanding of both phase boundaries may be within reach, although more experimental effort is required in this direction.

\textit{Acknowledgments.---} We thank A.-A. Haghighirad, M. Le Tacon, and S. Wilson for fruitful discussions. L.B., H.P. and T.B. were supported by the National Science Foundation through the University of Minnesota MRSEC under Award Number DMR-2011401. R.M.F. acknowledges support from the Mercator Fellowship from the German Research Foundation (DFG) through CRC TRR 288, 422213477 “Elasto-Q-Mat.”

\textit{Data availability.-} The data required to reproduce the results presented in this study is available at the Data Repository of the University of Minnesota~\cite{drum}.

\section*{End Matter}
\subsection*{Reciprocal lattice and high-symmetry wavevectors}
The primitive reciprocal lattice vectors of $P6/mmm$ (\#191) expressed in cartesian coordinates are 
\begin{equation}
    \mathbf{G}_1 = \frac{2\pi}{a}\begin{pmatrix} 1 \\ \frac{1}{\sqrt{3}} \\ 0 \end{pmatrix}, \quad
    \mathbf{G}_2 = \frac{2\pi}{a}\begin{pmatrix} 0 \\ \frac{2}{\sqrt{3}} \\ 0 \end{pmatrix}, \quad
    \mathbf{G}_3 = \frac{2\pi}{c}\begin{pmatrix} 0 \\ 0 \\ 1 \end{pmatrix}.
\end{equation}
where $a$ and $c$ are the lattice parameters. All wavevectors below, as well as in the main text, are given in reduced coordinates in the basis $(\mathbf{G}_1, \mathbf{G}_2, \mathbf{G}_3)$.

The three arms of the star of the zone boundary point $M$ are
\begin{equation}
    \mathbf{Q}_M^{(1)} = \left(\tfrac{1}{2},0,0\right), \quad
    \mathbf{Q}_M^{(2)} = \left(0,\tfrac{1}{2},0\right), \quad
    \mathbf{Q}_M^{(3)} = \left(\tfrac{1}{2},-\tfrac{1}{2},0\right),
\end{equation}
and the three arms of the star of $L$ are
\begin{equation}
    \mathbf{Q}_L^{(1)} = \left(\tfrac{1}{2},0,\tfrac{1}{2}\right), \quad
    \mathbf{Q}_L^{(2)} = \left(0,\tfrac{1}{2},\tfrac{1}{2}\right), \quad
    \mathbf{Q}_L^{(3)} = \left(\tfrac{1}{2},-\tfrac{1}{2},\tfrac{1}{2}\right).
\end{equation}
Three of the six arms of the star of $\Sigma$ are
\begin{equation}
    \mathbf{Q}_\Sigma^{(1)} = \left(\delta,0,0\right), \quad
    \mathbf{Q}_\Sigma^{(2)} = \left(0,\delta,0\right), \quad
    \mathbf{Q}_\Sigma^{(3)} = \left(\delta,-\delta,0\right),
\end{equation}
with time-reversal partners at $-\mathbf{Q}_\Sigma^{(i)}$, which are inequivalent under lattice translations. The complex order parameter $\sigma_i$ lives at $\mathbf{Q}_\Sigma^{(i)}$ with $\sigma_i^*$ at $-\mathbf{Q}_\Sigma^{(i)}$. Similarly, three out of six arms of the star of $R$ are
\begin{equation}
    \mathbf{Q}_R^{(1)} = \left(\delta,0,\tfrac{1}{2}\right), \quad
    \mathbf{Q}_R^{(2)} = \left(0,\delta,\tfrac{1}{2}\right), \quad
    \mathbf{Q}_R^{(3)} = \left(\delta,-\delta,\tfrac{1}{2}\right),
\end{equation}
with time-reversal partners at $-\mathbf{Q}_R^{(i)}$, and complex order parameter $r_i$ at $\mathbf{Q}_R^{(i)}$ with $r_i^*$ at $-\mathbf{Q}_R^{(i)}$. The zone-boundary points $M$ and $L$ have stars of size three because their wavevectors are invariant under inversion, whereas the $\Sigma$ and $R$ points have smaller little groups ($mm2$ rather than $mmm$), giving six arms in total. We note that the symmetry considerations above, including the little group $mm2$, the complex nature of the order parameters, and the trilinear couplings in the Landau free energy, apply to any point on the $\Sigma$ and $R$ lines with $\delta\in(0,1/2)$. In the main text we focused on $\delta=3/8$ since this is the commensurate value closest to the observed instability that is directly accessible on our $8\times8\times4$ $\mathbf{Q}$-point grid, and coincides with the experimentally reported ordering wavevector~\cite{Stier2024Dec}.

\subsection*{Fermi surface nesting}
To investigate the electronic contribution to the phonon instabilities, we calculate the Fermi surface nesting function
\begin{equation}
\xi(\mathbf{Q}) = \frac{1}{N_k}\sum_{\mathbf{k},n,m}
\delta(E_{n\mathbf{k}})\,\delta(E_{m,\mathbf{k+Q}})\,,
\label{eq:nesting}
\end{equation}
which reflects the purely geometric properties of the Fermi surace. It has been argued that a conventional Peierls instability requires concurrent sharp peaks in both the real and imaginary parts of the electronic susceptibility, causing all phonon modes at the nesting vector to soften~\cite{Johannes2008}. CsV$_3$Sb$_5$ clearly does not satisfy this criterion, as the instability is confined to a single phonon branch. Nevertheless, the calculated phonon spectrum inherits a momentum-dependent structure that closely tracks $\xi(\mathbf{Q})$. We therefore view Fermi surface nesting as a contributing factor rather than a complete mechanism for the instability.

\begin{figure}[h!]
    \centering
    \includegraphics[width=0.95\linewidth]{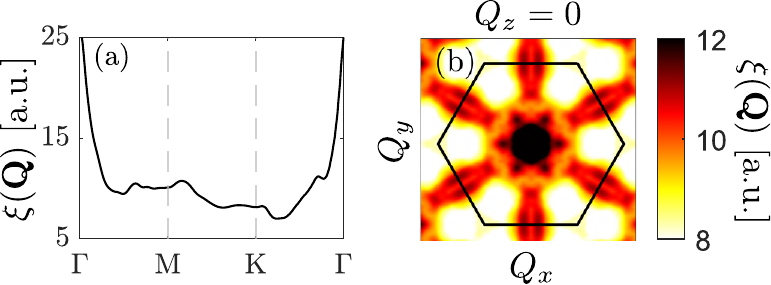}
    \caption{Fermi surface nesting function on the $Q_z=0$ plane. (a) Nesting on the high symmetry line $\Gamma$--M--K-$\Gamma$ line. (b) Nesting on the $Q_x$, $Q_y$ plane as a heat map. Contrary to $q_z=\pi/c$, the relative peaks close to the BZ edge are on the $\Gamma$--K line.}
    \label{fig:nesting}
\end{figure}

We constructed a 30-band Wannier model for CsV$_3$Sb$_5$ using \texttt{Wannier90}~\cite{Pizzi2020} and performed Wannier interpolation onto a fine reciprocal space grid. The nesting function $\xi(\mathbf{Q})$ was evaluated on a $1000\times 1000$ grid in $Q_x$, $Q_y$ at $Q_z = 0$ and $Q_z = \pi/c$. The delta functions were approximated by Gaussians of width $\sigma = 25$~meV. Results are shown in Fig.~\ref{fig:nesting} and Fig.~3(a–b).

\end{document}


\title{Supplemental Material - 3/8 charge-density wave instability in the kagome metal CsV$_3$Sb$_5$}

\author{Luca Buiarelli}
\thanks{These authors contributed equally to this work.}
\affiliation{Department of Chemical Engineering and Materials Science, University of Minnesota, MN 55455, USA}

\author{Hyeonseo H. Park}
\thanks{These authors contributed equally to this work.}
\affiliation{Department of Chemical Engineering and Materials Science, University of Minnesota, MN 55455, USA}

\author{Ethan T. Ritz}
\affiliation{Department of Engineering, Harvey Mudd College, Claremont, CA 91711, USA}
\affiliation{Department of Civil and Mineral Engineering, University of Toronto, Toronto, ON M5S 1A7, Canada}

\author{Rafael M. Fernandes}
\affiliation{Department of Physics, The Grainger College of Engineering, University of Illinois Urbana-Champaign, Urbana, IL 61801, USA}
\affiliation{Anthony J. Leggett Institute for Condensed Matter Theory, The Grainger College of Engineering, University of Illinois Urbana-Champaign, Urbana, IL 61801, USA}

\author{Turan Birol}
\thanks{Corresponding author: \href{mailto:tbirol@umn.edu}{tbirol@umn.edu}}
\affiliation{Department of Chemical Engineering and Materials Science, University of Minnesota, MN 55455, USA}

\date{\today}

\maketitle

\newcommand{\beginsupplement}{
  \setcounter{equation}{0}
  \renewcommand{\theequation}{S\arabic{equation}}
  \setcounter{figure}{0}
  \renewcommand{\thefigure}{S\arabic{figure}}
  \setcounter{table}{0}
  \renewcommand{\thetable}{S\arabic{table}}
}
\beginsupplement

\section{DFPT calculations}
We perform phonon calculations in CsV$_3$Sb$_5$ using density functional perturbation theory (DFPT) as implemented in Abinit~\cite{Gonze1997Apr, Torrent2008Apr, Audouze2008Jul, Gonze2020Mar}. We adopt the projector augmented waves (PAW)~\cite{Blochl1994Dec} formalism within the Perdew-Burke-Ernzerhof approximation of the exchange correlation functional for solids (PBEsol)~\cite{Perdew2008Apr}. We use a 650 eV energy cutoff for the plane waves basis and a 24$\times$24$\times$12 $\mathbf{k}$-point grid to calculate the electronic ground state properties, which we deem converged when the residue square of the wavefunctions is smaller than 10$^{-22}$ eV. We then use the perturbation theory to calculate the linear response to the phonons on a 8$\times$8$\times$4 $\mathbf{Q}$-point grid. The occupation was set by a gaussian smearing scheme with a smearing temperature of 1 meV. The phonons were calculated in the fully relaxed structure, where the forces applied on the atoms were minimized until they were smaller in modulus than 5 meV/\AA. The lattice parameters used were $a$=5.4414 \AA, $c$= 9.4308 \AA~and $z$=0.7395 for the free parameter of the apical Sb ions at the Wyckoff position $4h$.

\section{DFT supercell calculations}
The DFT calculations in the 8$\times$1$\times$2 (2$\times$1$\times$2) supercells were performed using the Vienna Ab-initio Simulation Package (VASP) version 6.5.0~\cite{vasp1,vasp2,vasp3}. We adopted the same density functional and smearing method with the DFPT calculations. Due to the large supercell sizes, we reduced the cutoff energy to 350 eV and used 2$\times$16$\times$4 (8$\times$16$\times$4) $\mathbf{k}$-point meshes. Before the calculations, the structures were relaxed with criteria of $10^{-7}$ eV for electronic energy difference and 1 meV/Å for residual atomic forces.

\begin{figure}
    \centering
    \includegraphics[width=0.4\linewidth]{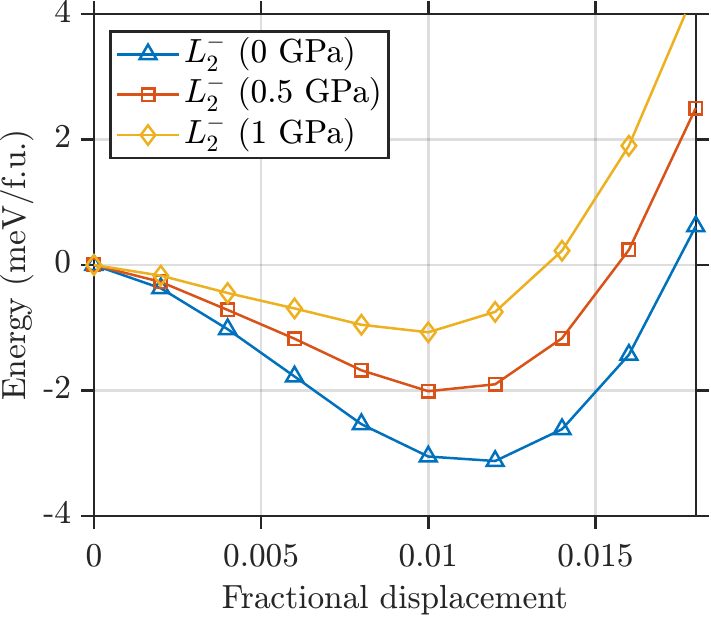}
    \includegraphics[width=0.4\linewidth]{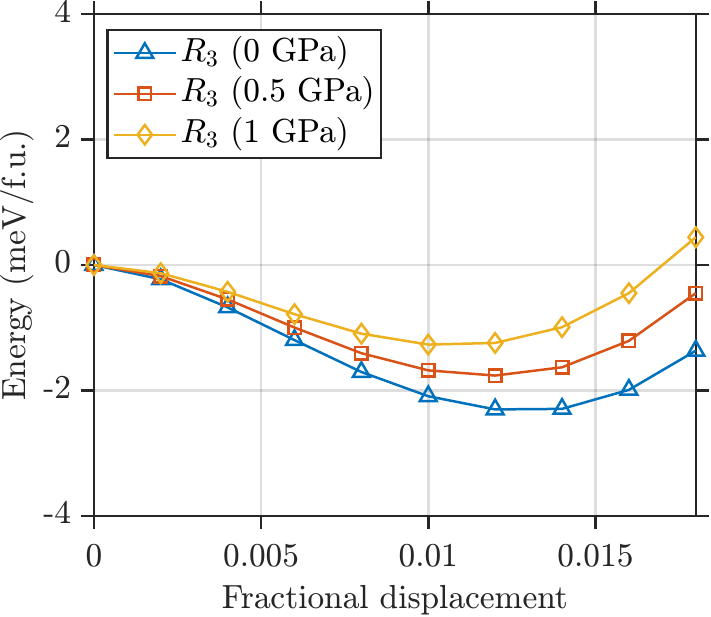}
    \caption{DFT energies obtained by distorting ions along the $L_2^-$ ($R_3$) phonon mode transforming as $l_1$ ($r_1$). The lattice parameters are fixed as the values at the origin. The fractional displacement on the x-axis is evaluated in the primitive unit cell using fractional coordinates.}
    \label{fig:energy_pressure}
\end{figure}

To obtain DFT energy curves as a function of displacements, we displaced ions along the eigenvectors of the dynamical matrix obtained in the DFPT calculations. 
The size of the displacements in Fig.~2(a) is defined as the root‑sum‑of‑squares displacement normalized by the square root of the supercell size relative to the primitive cell. In a similar way, we calculated enthalpy curves as a function of the same displacements in Fig.~2(b--c). In the enthalpy calculations, we fixed the displacements in fractional coordinate, relaxed lattice parameters, and added PV(pressure times volume) to the DFT energies. For comparision, we also evaluated DFT energies without relaxing unit cell but with increasing the displacements from the relaxed structure under pressure in Fig.~\ref{fig:energy_pressure}. The energy behavior here is qualitatively the same as that in Fig.~2(b--c).

To analyze the band reconstruction associated with different phonon modes, we displaced the ions along the eigenvectors obtained from DFPT and obtained the corresponding pseudo-wavefunctions from VASP. The resulting supercell band structure was then unfolded onto the Brillouin zone of the primitive cell. The unfolding was performed using the \texttt{VaspBandUnfolding} code~\cite{ZhengVaspBandUnfolding}, following the spectral-weight formalism of Popescu and Zunger~\cite{Popescu2012}. Specifically, the plane-wave coefficients of the supercell pseudo-wavefunctions were used to determine the spectral weight at the corresponding primitive wave vector, thereby recovering the effective band structure in the primitive Brillouin zone.

\section{Lack of dependence on the phase of single-$\mathbf{Q}$ order parameters}
\begin{figure}
    \centering
    \includegraphics[width=0.4\linewidth]{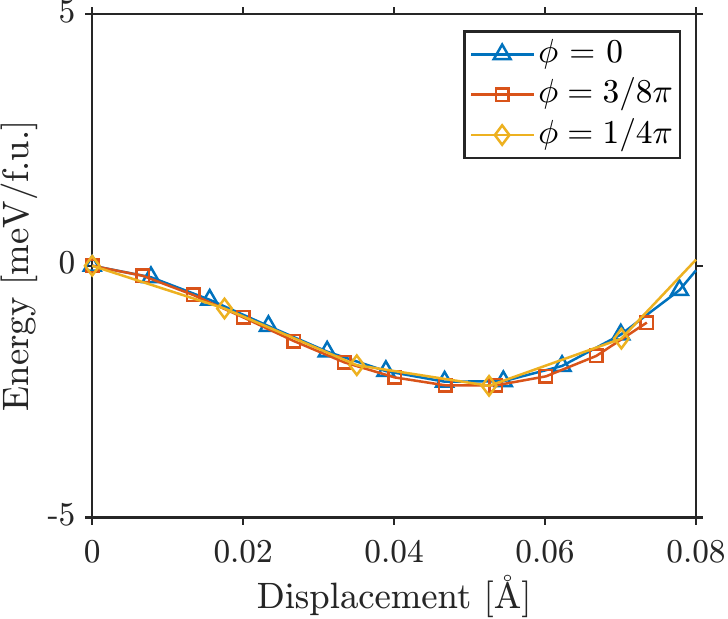}
    \caption{DFT energies obtained by distorting ions along the ($R_3$) phonon mode with different phases $\phi$.}
    \label{fig:phase_independence}
\end{figure}

In the main text, we define $\mathbf{R} = (r_1, r_2, r_3)$, where $r_n = |r_n|\,e^{i\phi}$ is a complex order parameter. We set $\phi = 0$ when the distorted structure has $Cmmm$ (\#65) symmetry, consistent with the convention of the \textsc{Isosubgroup} tool in the ISOTROPY Software Suite~\cite{Isotropy}. For commensurate order at the wave vector $\mathbf{Q}_R^{(1)} = \left(\frac{3}{8}, 0, \frac{1}{2}\right)$, the \textsc{Invariants} tool of the ISOTROPY Software Suite~\cite{Isotropy} shows that phase-dependent terms in a free energy containing only a single component $r_i$ first appear at eighth order. Such terms are not expected to have a strong impact on the energy. To verify that the DFT energy is indeed phase independent, we computed it along several values of $\phi$: $\phi = 0$ ($Cmmm$, \#65), $\phi = \pi/4$ ($Amm2$, \#38), and $\phi = 3\pi/8$ ($Cmcm$, \#63). The results, shown in Fig.~\ref{fig:phase_independence}, confirm that the energy does not depend on $\phi$, as expected: for incommensurate order this independence holds to all orders.

\section{Fourth-order Free energy Expansion}
Using the order parameters introduced in the main text, we construct all symmetry-allowed invariants up to fourth order in the space group $P6/mmm$, as obtained from the \textsc{Invariants} tool of the ISOTROPY Software Suite~\cite{Isotropy}. 
All second and third-order terms were discussed in the main text; here, we list the fourth-order invariants, which capture the couplings between $M_1^+$, $L_2^-$, $\Sigma_1$, and $R_3$ order parameters.

We begin with the “trivial” fourth-order terms, which depend on a single order parameter:
\begin{equation}
\begin{split}
    \mathcal{F}_M^{(4)} = \frac{u_M}{4}|\mathbf{M}|^4 + \frac{\lambda_M}{4}(m_1^2m_2^2+m_2^2m_3^2+m_1^2m_3^2),
\end{split}
\end{equation}

\begin{equation}
\begin{split}
    \mathcal{F}_L^{(4)} = \frac{u_L}{4}|\mathbf{L}|^4 + \frac{\lambda_L}{4}(l_1^2l_2^2+l_2^2l_3^2+l_1^2l_3^2),
\end{split}
\end{equation}

\begin{equation}
\mathcal{F}_{\Sigma}^{(4)} = \frac{u_\Sigma}{4}|\boldsymbol{\Sigma}|^4 + \frac{\lambda_\Sigma}{4}(|\sigma_1|^2|\sigma_2|^2 + |\sigma_2|^2|\sigma_3|^2 + |\sigma_3|^2|\sigma_1|^2),
\end{equation}
\begin{equation}
\mathcal{F}_{R}^{(4)} = \frac{u_R}{4}|\mathbf{R}|^4 
+ \frac{\lambda_R}{4}(|r_1|^2|r_2|^2 + |r_2|^2|r_3|^2 + |r_3|^2|r_1|^2),
\end{equation}
where $|\mathbf{X}|^4=(|x_1|^2+|x_2|^2+|x_3|^2)^2$, for $\mathbf{X}=\mathbf{M},\mathbf{L},\mathbf{\Sigma},\mathbf{R}$.

Next, we list the mixed fourth-order invariants coupling $M_1^+$ and $L_2^-$, previously discussed in Refs.~\cite{Christensen2021Dec, Ritz2023May}:
\begin{equation}
\begin{split}
    \mathcal{F}_{ML}^{(4)} = \frac{\lambda^{(1)}_{ML}}{4}(m_1m_2l_1l_2+m_2m_3l_2l_3+m_1m_3l_1l_3) + \frac{\lambda^{(2)}_{ML}}{4}(m_1^2l_1^2+m_2^2l_2^2+m_3^2l_3^2) + \frac{\lambda^{(3)}_{ML}}{4}|\mathbf{M}|^2|\mathbf{L}|^2.
\end{split}
\end{equation}

Analogously, a set of more intricate fourth-order terms couple $\Sigma_1$ and $R_3$:
\begin{equation}
\begin{split}
\mathcal{F}_{\Sigma R}^{(4)} &= \frac{\lambda^{(1)}_{\Sigma R}}{4}|\boldsymbol{\Sigma}|^2|\mathbf{R}|^2 + \frac{\lambda^{(2)}_{\Sigma R}}{2}\sum_{i=1}^3|r_i|^2|\sigma_i|^2 + \frac{\lambda^{(3)}_{\Sigma R}}{4}\sum_{i=1}^3\left[\mathrm{Re}(r_i\sigma_i^*)\right]^2 \\
&+ \frac{\lambda^{(4)}_{\Sigma R}}{4}\sum_{i\neq j}^3\mathrm{Re}(r_i\sigma_i^*)\mathrm{Re}(r_j\sigma_j^*) + \frac{\lambda^{(5)}_{\Sigma R}}{4}\sum_{k=1}^3 \epsilon_{ijk} \mathrm{Im}(r_i\sigma_i^*)\mathrm{Im}(r_j\sigma_j^*),
\end{split}
\end{equation} where $\epsilon_{ijk}$ is a Levi-Civita symbol.

Finally, the fourth-order terms that couple all four order parameters read
\begin{equation}
\mathcal{F}_{ML\Sigma R}^{(4)} = \frac{\lambda^{(1)}_{ML\Sigma R}}{4}
\left(\sum_{i=1}^3 m_i l_i\right)\left(\sum_{i=1}^3\mathrm{Re}(\sigma_i r_i^*)\right)
+ \frac{\lambda^{(2)}_{ML\Sigma R}}{4}\sum_{i\neq j}^3 m_i l_i\,\mathrm{Re}(\sigma_j r_j^*),
\label{eq:ql}
\end{equation}
along with additional pairwise coupling terms,
\begin{equation}
\mathcal{F}_{M\Sigma}^{(4)} = \frac{\lambda^{(1)}_{M\Sigma}}{4}|\mathbf{M}|^2|\boldsymbol{\Sigma}|^2 
+ \frac{\lambda^{(2)}_{M\Sigma}}{4}\sum_{i=1}^3 m_i^2|\sigma_i|^2
\label{eq:bq1}
\end{equation}

\begin{equation}
\mathcal{F}_{MR}^{(4)} = \frac{\lambda^{(1)}_{MR}}{4}|\mathbf{M}|^2|\mathbf{R}|^2 
+ \frac{\lambda^{(2)}_{MR}}{4}\sum_{i=1}^3 m_i^2|r_i|^2
\label{eq:bq2}
\end{equation}

\begin{equation}
\mathcal{F}_{L\Sigma}^{(4)} = \frac{\lambda^{(1)}_{L\Sigma}}{4}|\mathbf{L}|^2|\boldsymbol{\Sigma}|^2 
+ \frac{\lambda^{(2)}_{L\Sigma}}{4}\sum_{i=1}^3 l_i^2|\sigma_i|^2
\label{eq:bq3}
\end{equation}

\begin{equation}
\mathcal{F}_{LR}^{(4)} = \frac{\lambda^{(1)}_{LR}}{4}|\mathbf{L}|^2|\mathbf{R}|^2 
+ \frac{\lambda^{(2)}_{LR}}{4}\sum_{i=1}^3 l_i^2|r_i|^2
\label{eq:bq4}
\end{equation}
Together, these terms form the complete set of fourth-order invariants coupling $\mathbf{M}$, $\mathbf{L}$, $\boldsymbol{\Sigma}$, and $\mathbf{R}$ in $P6/mmm$ symmetry.

In the main text, we refer to the terms in Eq.~\ref{eq:ql} as quadrilinear and those in Eqs.~\ref{eq:bq1}--\ref{eq:bq4} as biquadratic. Quadrilinear terms can always lower the energy by adjusting the signs of the order parameters, and therefore favor their coexistence. Biquadratic terms, by contrast, generally raise the energy when different order parameters coexist, assuming positive coefficients. Which of these dominates determines whether the order parameters condense together or separately. When the quadrilinear coupling is strong relative to the biquadratic one, the order parameters prefer to coexist, giving rise to phases with multiple order parameters. In the opposite limit, the system favors condensing either the zone-boundary modes ($M_1^+$ and $L_2^-$) or the high-symmetry-line modes ($\Sigma_1$ and $R_3$) independently, and one may observe domains with different condensates or a single domain with one type of condensate.

\section{Candidate ground states}
Using the ISOSUBGROUP tool in the ISOTROPY Software Suite~\cite{Isotropy}, we explore the possible ground states that are allowed by the free energy coupling the order parameters $\mathbf{M}$, $\mathbf{L}$, $\mathbf{\Sigma}$ and $\mathbf{R}$. The results are shown in Tables \ref{tab:subgroup_r}, \ref{tab:subgroup_sr}, \ref{tab:subgroup_mlr} and \ref{tab:subgroup_mlsr}.

\begin{table}[]
\centering
\begin{tabular}{lr}
\hline
OP $\mathbf{R}$ Direction  & Space Group \\
\hline
$\Big(a,0;a,0;a,0\Big)$ & $P6/mmm$ (\#191) \\
$\Big(a,0;a,0;0,a\Big)$ & $P6_3/mmc$ (\#194) \\
$\Big(a,(\sqrt{2}-1)a;0,0;0,0\Big)$ & $Cmcm$ (\#63) \\
$\Big(a,0;0,0;0,0\Big)$ & $Cmmm$ (\#65) \\
$\Big(a,0;a,0;0,0\Big)$ & $Fmmm$ (\#69) \\
$\Big(a,(\sqrt{2}+1)a;a,(\sqrt{2}+1)a;0,0\Big)$ & $Fmmm$ (\#69) \\
$\Big(a,b;0,0;0,0\Big)$ & $Amm2$ (\#38) \\
$\Big(a,b;b,a;a,b\Big)$ & $P\bar{6}m2$ (\#187) \\
$\Big(a,b;b,a;0,0\Big)$ & $Fmm2$ (\#42) \\
$\Big(a,0;b,0;0,0\Big)$ & $C2/m$ (\#12) \\
$\Big(a,(\sqrt{2}+1)a;b,(\sqrt{2}+1)b;0,0\Big)$ & $C2/m$ (\#12) \\
$\Big(a,(\sqrt{2}-1)a;b,0;0,0\Big)$ & $C2/m$ (\#12) \\
$\Big(a,b;a,b;0,0\Big)$ & $Fmm2$ (\#42) \\
$\Big(a,0;b,0;a,0\Big)$ & $Cmmm$ (\#65) \\
$\Big(a,0;0,b;a,0\Big)$ & $Cmcm$ (\#63) \\
$\Big(a,(\sqrt{2}+1)a;a,(\sqrt{2}+1)a;b,0\Big)$ & $Cmmm$ (\#65) \\
$\Big(a,(\sqrt{2}-1)a;a,(\sqrt{2}-1)a;0,b\Big)$ & $Cmcm$ (\#63) \\
$\Big(a,0;b,0;c,0\Big)$ & $P2/m$ (\#10) \\
$\Big(a,0;0,b;c,0\Big)$ & $P2_1/m$ (\#11) \\
$\Big(a,(\sqrt{2}+1)a;b,(\sqrt{2}+1)b;c,0\Big)$ & $P2/m$ (\#10) \\
$\Big(a,(\sqrt{2}-1)a;b,(\sqrt{2}-1)b;0,c\Big)$ & $P2_1/m$ (\#11) \\
$\Big(a,b;a,b;0,c\Big)$ & $Ama2$ (\#40) \\
$\Big(a,b;a,b;c,0\Big)$ & $Amm2$ (\#38) \\
$\Big(a,b;c,d;0,0\Big)$ & $Cm$ (\#8) \\
$\Big(a,b;c,d;d,c\Big)$ & $Amm2$ (\#38) \\
$\Big(a,b;c,d;e,f\Big)$ & $Pm$ (\#6) \\
\hline
\end{tabular}
\caption{Candidate ground states for the free energy containing only terms in the order parameter $\mathbf{R}$. We use the notation $\mathbf{R}(a,b;c,d;e,f)$ for $r_1=a+ib$, $r_2=c+id$, $r_3=e+if$, consistent with the complex order parameter $\mathbf{R} = (r_1,r_2,r_3)$ defined in the main text.}
\label{tab:subgroup_r}
\end{table}

\begin{table}[]
\centering
\begin{tabular}{llr}
\hline
OP $\mathbf{\Sigma}$ direction & OP $\mathbf{R}$ direction & Space Group \\
\hline
$\Big(a,0;a,0;a,0\Big)$ & $\Big(b,0;b,0;b,0\Big)$ & $P6/mmm$ (\#191) \\
$\Big(a,0;a,0;a,0\Big)$ & $\Big(0,-b;0,b;0,-b\Big)$ & $P6_3/mmc$ (\#194) \\
$\Big(a,(\sqrt{2}+1)a;0,0;0,0\Big)$ & $\Big(-(\sqrt{2}-1)b,-b;0,0;0,0\Big)$ & $Pmma$ (\#51) \\
$\Big(a,(\sqrt{2}+1)a;0,0;0,0\Big)$ & $\Big(b,-(\sqrt{2}-1)b;0,0;0,0\Big)$ & $Pmmn$ (\#59) \\
$\Big(a,0;0,0;0,0\Big)$ & $\Big(b,0;0,0;0,0\Big)$ & $Pmmm$ (\#47) \\
$\Big(a,0;0,0;0,0\Big)$ & $\Big(0,b;0,0;0,0\Big)$ & $Pmma$ (\#51) \\
$\Big(a,0;0,0;0,0\Big)$ & $\Big(0,0;b,0;b,0\Big)$ & $Fmmm$ (\#69) \\
$\Big(a,0;b,0;a,0\Big)$ & $\Big(c,0;0,0;-c,0\Big)$ & $Cccm$ (\#66) \\
$\Big(a,0;b,0;a,0\Big)$ & $\Big(0,-c;0,0;0,c\Big)$ & $Cmcm$ (\#63) \\
$\Big(a,0;0,0;0,0\Big)$ & $\Big(0,0;b,(\sqrt{2}+1)b;b,(\sqrt{2}+1)b\Big)$ & $Fmmm$ (\#69) \\
$\Big(a,(\sqrt{2}+1)a;a,(\sqrt{2}+1)a;b,0\Big)$ & $\Big(-c,-(\sqrt{2}+1)c;c,(\sqrt{2}+1)c;0,0\Big)$ & $Cccm$ (\#66) \\
$\Big(a,(\sqrt{2}+1)a;a,(\sqrt{2}+1)a;b,0\Big)$ & $\Big(-(\sqrt{2}+1)c,c;-(\sqrt{2}+1)c,c;0,0\Big)$ & $Cmcm$ (\#63) \\
$\Big(a,b;0,0;0,0\Big)$ & $\Big(c,d;0,0;0,0\Big)$ & $Pmm2$ (\#25) \\
$\Big(a,b;b,a;a,b\Big)$ & $\Big(c,d;d,c;c,d\Big)$ & $P\text{-}6m2$ (\#187) \\
$\Big(a,b;0,0;0,0\Big)$ & $\Big(0,0;c,d;d,c\Big)$ & $Fmm2$ (\#42) \\
$\Big(a,b;c,d;d,c\Big)$ & $\Big(0,0;e,-f;f,-e\Big)$ & $Ama2$ (\#40) \\
$\Big(a,0;0,0;0,0\Big)$ & $\Big(0,0;b,0;c,0\Big)$ & $C2/m$ (\#12) \\
$\Big(a,0;0,0;0,0\Big)$ & $\Big(0,0;b,(\sqrt{2}+1)b;c,(\sqrt{2}+1)c\Big)$ & $C2/m$ (\#12) \\
$\Big(a,(\sqrt{2}+1)a;0,0;0,0\Big)$ & $\Big(0,0;-(\sqrt{2}-1)b,-b;c,0\Big)$ & $C2/m$ (\#12) \\
$\Big(a,0;0,0;0,0\Big)$ & $\Big(0,0;b,c;b,c\Big)$ & $Fmm2$ (\#42) \\
$\Big(a,0;b,0;a,0\Big)$ & $\Big(c,0;d,0;c,0\Big)$ & $Cmmm$ (\#65) \\
$\Big(a,0;b,0;a,0\Big)$ & $\Big(0,-c;0,-d;0,-c\Big)$ & $Cmcm$ (\#63) \\
$\Big(a,(\sqrt{2}+1)a;a,(\sqrt{2}+1)a;b,0\Big)$ & $\Big(c,(\sqrt{2}+1)c;c,(\sqrt{2}+1)c;d,0\Big)$ & $Cmmm$ (\#65) \\
$\Big(a,(\sqrt{2}+1)a;a,(\sqrt{2}+1)a;b,0\Big)$ & $\Big(c,-(\sqrt{2}-1)c;-c,(\sqrt{2}-1)c;0,-d\Big)$ & $Cmcm$ (\#63) \\
$\Big(a,0;b,0;c,0\Big)$ & $\Big(d,0;e,0;f,0\Big)$ & $P2/m$ (\#10) \\
$\Big(a,0;b,0;c,0\Big)$ & $\Big(0,-d;0,e;0,f\Big)$ & $P2_1/m$ (\#11) \\
$\Big(a,(\sqrt{2}+1)a;b,(\sqrt{2}+1)b;c,0\Big)$ & $\Big(d,(\sqrt{2}+1)d;e,(\sqrt{2}+1)e;f,0\Big)$ & $P2/m$ (\#10) \\
$\Big(a,(\sqrt{2}+1)a;b,(\sqrt{2}+1)b;c,0\Big)$ & $\Big(-d,(\sqrt{2}-1)d;e,-(\sqrt{2}-1)e;0,-f\Big)$ & $P2_1/m$ (\#11) \\
$\Big(a,b;a,b;c,0\Big)$ & $\Big(d,e;-d,-e;0,-f\Big)$ & $Ama2$ (\#40) \\
$\Big(a,b;a,b;c,0\Big)$ & $\Big(d,e;d,e;f,0\Big)$ & $Amm2$ (\#38) \\
$\Big(a,b;0,0;0,0\Big)$ & $\Big(0,0;c,d;e,f\Big)$ & $Cm$ (\#8) \\
$\Big(a,b;c,d;d,c\Big)$ & $\Big(e,f;g,h;h,g\Big)$ & $Amm2$ (\#38) \\
$\Big(a,b;c,d;e,f\Big)$ & $\Big(g,h;i,j;k,l\Big)$ & $Pm$ (\#6) \\
\hline
\end{tabular}
\caption{Candidate ground states for the free energy containing only terms in the order paramters $\mathbf{\Sigma}$ and $\mathbf{R}$. We use the notation $\mathbf{R}(a,b;c,d;e,f)$ for $r_1=a+ib$, $r_2=c+id$, $r_3=e+if$, and $\mathbf{\Sigma}(a,b;c,d;e,f)$ for $\sigma_1=a+ib$, $\sigma_2=c+id$, $\sigma_3=e+if$, consistent with the complex order parameters $\mathbf{R} = (r_1,r_2,r_3)$ and $\mathbf{\Sigma} = (\sigma_1,\sigma_2,\sigma_3)$ defined in the main text.}
\label{tab:subgroup_sr}
\end{table}

\begin{table}[]
\centering
\begin{tabular}{lllr}
\hline
OP $\mathbf{M}$ & OP $\mathbf{L}$ & OP $\mathbf{R}$ & Space Group \\
\hline
$\Big(a;a;a\Big)$ & $\Big(b;b;b\Big)$ & $\Big(c,0;c,0;c,0\Big)$  & $P6/mmm$ (\#191) \\
$\Big(a;0;0\Big)$ & $\Big(b;0;0\Big)$ & $\Big(0,0;c,0.414c;0,0\Big)$  & $P2/m$ (\#10) \\
$\Big(a;0;0\Big)$ & $\Big(0;b;c\Big)$ & $\Big(0,0;d,0.414d;0,0\Big)$  & $P2_1/m$ (\#11) \\
$\Big(a;0;0\Big)$ & $\Big(b;0;0\Big)$ & $\Big(c,0;0,0;0,0\Big)$  & $Pmmm$ (\#47) \\
$\Big(a;0;0\Big)$ & $\Big(0;b;b\Big)$ & $\Big(c,0;0,0;0,0\Big)$  & $Immm$ (\#71) \\
$\Big(a;0;0\Big)$ & $\Big(0;b;b\Big)$ & $\Big(-0.707c,-0.707c;0,0;0,0\Big)$  & $Imma$ (\#74) \\
$\Big(a;b;a\Big)$ & $\Big(-c;0;c\Big)$ & $\Big(0,0;d,0;0,0\Big)$  & $Pmna$ (\#53) \\
$\Big(a;b;a\Big)$ & $\Big(-c;0;c\Big)$ & $\Big(0,0;-0.707d,-0.707d;0,0\Big)$  & $Pnnm$ (\#58) \\
$\Big(a;0;0\Big)$ & $\Big(0;b;c\Big)$ & $\Big(d,0;0,0;0,0\Big)$  & $C2/m$ (\#12) \\
$\Big(a;b;a\Big)$ & $\Big(c;d;c\Big)$ & $\Big(0,0;e,0;0,0\Big)$  & $Pmmm$ (\#47) \\
$\Big(a;b;a\Big)$ & $\Big(c;d;c\Big)$ & $\Big(0,0;-0.707e,-0.707e;0,0\Big)$  & $Pmma$ (\#51) \\
$\Big(a;b;c\Big)$ & $\Big(d;e;f\Big)$ & $\Big(g,0;0,0;0,0\Big)$  & $P2/m$ (\#10) \\
$\Big(a;b;a\Big)$ & $\Big(-c;0;c\Big)$ & $\Big(d,0;0,0;d,0\Big)$  & $Cccm$ (\#66) \\
$\Big(a;0;0\Big)$ & $\Big(0;b;b\Big)$ & $\Big(0,0;c,2.414c;c,2.414c\Big)$  & $Cmcm$ (\#63) \\
$\Big(a;0;0\Big)$ & $\Big(b;0;0\Big)$ & $\Big(c,d;0,0;0,0\Big)$  & $Pmm2$ (\#25) \\
$\Big(a;0;0\Big)$ & $\Big(0;b;b\Big)$ & $\Big(c,d;0,0;0,0\Big)$  & $Imm2$ (\#44) \\
$\Big(a;b;a\Big)$ & $\Big(-c;0;c\Big)$ & $\Big(0,0;d,e;0,0\Big)$  & $Pmn2_1$ (\#31) \\
$\Big(a;0;0\Big)$ & $\Big(0;b;c\Big)$ & $\Big(d,e;0,0;0,0\Big)$  & $Cm$ (\#8) \\
$\Big(a;b;a\Big)$ & $\Big(c;d;c\Big)$ & $\Big(0,0;e,f;0,0\Big)$  & $Pmm2$ (\#25) \\
$\Big(a;b;c\Big)$ & $\Big(d;e;f\Big)$ & $\Big(g,h;0,0;0,0\Big)$  & $Pm$ (\#6) \\
$\Big(a;a;a\Big)$ & $\Big(b;b;b\Big)$ & $\Big(c,d;d,c;c,d\Big)$  & $P\bar{6}m2$ (\#187) \\
$\Big(a;b;a\Big)$ & $\Big(-c;0;c\Big)$ & $\Big(e,-d;0,0;d,e\Big)$  & $Ama2$ (\#40) \\
$\Big(a;b;a\Big)$ & $\Big(c;d;c\Big)$ & $\Big(e,0;f,0;e,0\Big)$  & $Cmmm$ (\#65) \\
$\Big(a;0;0\Big)$ & $\Big(b;0;0\Big)$ & $\Big(d,0;c,2.414c;c,2.414c\Big)$  & $Cmmm$ (\#65) \\
$\Big(a;0;0\Big)$ & $\Big(0;b;b\Big)$ & $\Big(0.707d,0.707d;c,0.414c;-0.414c,-c\Big)$  & $Cmcm$ (\#63) \\
$\Big(a;b;c\Big)$ & $\Big(d;e;f\Big)$ & $\Big(g,0;h,0;i,0\Big)$  & $P2/m$ (\#10) \\
$\Big(a;0;0\Big)$ & $\Big(b;0;0\Big)$ & $\Big(e,0;c,2.414c;d,2.414d\Big)$  & $P2/m$ (\#10) \\
$\Big(a;0;0\Big)$ & $\Big(0;b;c\Big)$ & $\Big(0,-f;d,0.414d;e,0.414e\Big)$  & $P2_1/m$ (\#11) \\
$\Big(a;b;a\Big)$ & $\Big(-c;0;c\Big)$ & $\Big(d,-e;0,-f;d,e\Big)$  & $Ama2$ (\#40) \\
$\Big(a;b;a\Big)$ & $\Big(c;d;c\Big)$ & $\Big(e,-f;g,0;e,f\Big)$  & $Amm2$ (\#38) \\
$\Big(a;b;a\Big)$ & $\Big(c;d;c\Big)$ & $\Big(h,-g;e,f;g,h\Big)$  & $Amm2$ (\#38) \\
$\Big(a;b;c\Big)$ & $\Big(d;e;f\Big)$ & $\Big(g,h;i,j;k,l\Big)$  & $Pm$ (\#6) \\
\hline
\end{tabular}
\caption{Candidate ground states for the free energy containing only terms in the order parameters $\mathbf{M}$, $\mathbf{L}$, and $\mathbf{R}$. We use the notation $\mathbf{M}(a,b,c)$ for $m_1=a$, $m_2=b$, $m_3=c$; $\mathbf{L}(a,b,c)$ for $l_1=a$, $l_2=b$, $l_3=c$; and $\mathbf{R}(a,b;c,d;e,f)$ for $r_1=a+ib$, $r_2=c+id$, $r_3=e+if$, consistent with the real order parameters $\mathbf{M} = (m_1,m_2,m_3)$ and $\mathbf{L} = (l_1,l_2,l_3)$, and the complex order parameter $\mathbf{R} = (r_1,r_2,r_3)$, defined in the main text.}
\label{tab:subgroup_mlr}
\end{table}

\begin{table}[]
\centering
\begin{tabular}{llllr}
\hline
OP $\mathbf{M}$ & OP $\mathbf{L}$ & OP $\mathbf{\Sigma}$ & OP $\mathbf{R}$ & Space Group \\
\hline
$\Big(a;a;a\Big)$ & $\Big(b;b;b\Big)$ & $\Big(c,0;c,0;c,0\Big)$ & $\Big(d,0;d,0;d,0\Big)$ & $P6/mmm$ (\#191) \\
$\Big(a;0;0\Big)$ & $\Big(b;0;0\Big)$ & $\Big(0,0;c,(\sqrt{2}+1)c;0,0\Big)$ & $\Big(0,0;-(\sqrt{2}-1)d,-d;0,0\Big)$ & $P2/m$ (\#10) \\
$\Big(a;0;0\Big)$ & $\Big(0;b;c\Big)$ & $\Big(0,0;d,(\sqrt{2}+1)d;0,0\Big)$ & $\Big(0,0;e,-(\sqrt{2}-1)e;0,0\Big)$ & $P2_1/m$ (\#11) \\
$\Big(a;0;0\Big)$ & $\Big(b;0;0\Big)$ & $\Big(c,0;0,0;0,0\Big)$ & $\Big(d,0;0,0;0,0\Big)$ & $Pmmm$ (\#47) \\
$\Big(a;0;0\Big)$ & $\Big(0;b;b\Big)$ & $\Big(c,0;0,0;0,0\Big)$ & $\Big(0,d;0,0;0,0\Big)$ & $Pmmn$ (\#59) \\
$\Big(a;0;0\Big)$ & $\Big(0;b;b\Big)$ & $\Big(-c/\sqrt{2},-c/\sqrt{2};0,0;0,0\Big)$ & $\Big(d/\sqrt{2},-d/\sqrt{2};0,0;0,0\Big)$ & $Pmma$ (\#51) \\
$\Big(a;b;a\Big)$ & $\Big(-c;0;c\Big)$ & $\Big(0,0;d,0;0,0\Big)$ & $\Big(0,0;0,e;0,0\Big)$ & $Pma2$ (\#28) \\
$\Big(a;b;a\Big)$ & $\Big(-c;0;c\Big)$ & $\Big(0,0;-d/\sqrt{2},-d/\sqrt{2};0,0\Big)$ & $\Big(0,0;e/\sqrt{2},-e/\sqrt{2};0,0\Big)$ & $Pmn2_1$ (\#31) \\
$\Big(a;0;0\Big)$ & $\Big(0;b;c\Big)$ & $\Big(d,0;0,0;0,0\Big)$ & $\Big(0,e;0,0;0,0\Big)$ & $P2_1/m$ (\#11) \\
$\Big(a;b;a\Big)$ & $\Big(c;d;c\Big)$ & $\Big(0,0;e,0;0,0\Big)$ & $\Big(0,0;f,0;0,0\Big)$ & $Pmmm$ (\#47) \\
$\Big(a;b;a\Big)$ & $\Big(c;d;c\Big)$ & $\Big(0,0;-e/\sqrt{2},-e/\sqrt{2};0,0\Big)$ & $\Big(0,0;-f/\sqrt{2},-f/\sqrt{2};0,0\Big)$ & $Pmma$ (\#51) \\
$\Big(a;b;c\Big)$ & $\Big(d;e;f\Big)$ & $\Big(g,0;0,0;0,0\Big)$ & $\Big(h,0;0,0;0,0\Big)$ & $P2/m$ (\#10) \\
$\Big(a;b;a\Big)$ & $\Big(-c;0;c\Big)$ & $\Big(d,0;e,0;d,0\Big)$ & $\Big(f,0;0,0;-f,0\Big)$ & $Cccm$ (\#66) \\
$\Big(a;0;0\Big)$ & $\Big(0;b;b\Big)$ & $\Big(d,0;c,(\sqrt{2}+1)c;c,(\sqrt{2}+1)c\Big)$ & $\Big(0,0;-(\sqrt{2}+1)e,e;-(\sqrt{2}+1)e,e\Big)$ & $Cmcm$ (\#63) \\
$\Big(a;0;0\Big)$ & $\Big(b;0;0\Big)$ & $\Big(c,d;0,0;0,0\Big)$ & $\Big(e,f;0,0;0,0\Big)$ & $Pmm2$ (\#25) \\
$\Big(a;b;a\Big)$ & $\Big(c;d;c\Big)$ & $\Big(0,0;e,f;0,0\Big)$ & $\Big(0,0;g,h;0,0\Big)$ & $Pmm2$ (\#25) \\
$\Big(a;b;c\Big)$ & $\Big(d;e;f\Big)$ & $\Big(g,h;0,0;0,0\Big)$ & $\Big(i,j;0,0;0,0\Big)$ & $Pm$ (\#6) \\
$\Big(a;a;a\Big)$ & $\Big(b;b;b\Big)$ & $\Big(c,d;d,c;c,d\Big)$ & $\Big(e,f;f,e;e,f\Big)$ & $P\bar{6}m2$ (\#187) \\
$\Big(a;b;a\Big)$ & $\Big(-c;0;c\Big)$ & $\Big(g,-f;d,e;f,g\Big)$ & $\Big(i,h;0,0;h,-i\Big)$ & $Ama2$ (\#40) \\
$\Big(a;b;a\Big)$ & $\Big(c;d;c\Big)$ & $\Big(e,0;f,0;e,0\Big)$ & $\Big(g,0;h,0;g,0\Big)$ & $Cmmm$ (\#65) \\
$\Big(a;0;0\Big)$ & $\Big(b;0;0\Big)$ & $\Big(d,0;c,(\sqrt{2}+1)c;c,(\sqrt{2}+1)c\Big)$ & $\Big(f,0;e,(\sqrt{2}+1)e;e,(\sqrt{2}+1)e\Big)$ & $Cmmm$ (\#65) \\
$\Big(a;0;0\Big)$ & $\Big(0;b;b\Big)$ & $\Big(-d/\sqrt{2},d/\sqrt{2};c,(\sqrt{2}+1)c;c,-(\sqrt{2}+1)c\Big)$ & $\Big(-f/\sqrt{2},-f/\sqrt{2};e,-(\sqrt{2}-1)e;e,(\sqrt{2}-1)e\Big)$ & $Cmcm$ (\#63) \\
$\Big(a;b;c\Big)$ & $\Big(d;e;f\Big)$ & $\Big(g,0;h,0;i,0\Big)$ & $\Big(j,0;k,0;l,0\Big)$ & $P2/m$ (\#10) \\
$\Big(a;0;0\Big)$ & $\Big(b;0;0\Big)$ & $\Big(e,0;c,(\sqrt{2}+1)c;d,(\sqrt{2}+1)d\Big)$ & $\Big(h,0;f,(\sqrt{2}+1)f;g,(\sqrt{2}+1)g\Big)$ & $P2/m$ (\#10) \\
$\Big(a;0;0\Big)$ & $\Big(0;b;c\Big)$ & $\Big(f,0;d,(\sqrt{2}+1)d;e,(\sqrt{2}+1)e\Big)$ & $\Big(0,i;-h,(\sqrt{2}-1)h;-g,(\sqrt{2}-1)g\Big)$ & $P2_1/m$ (\#11) \\
$\Big(a;b;a\Big)$ & $\Big(-c;0;c\Big)$ & $\Big(d,-e;f,0;d,e\Big)$ & $\Big(-g,h;0,i;g,h\Big)$ & $Ama2$ (\#40) \\
$\Big(a;b;a\Big)$ & $\Big(c;d;c\Big)$ & $\Big(e,-f;g,0;e,f\Big)$ & $\Big(h,-i;j,0;h,i\Big)$ & $Amm2$ (\#38) \\
$\Big(a;b;a\Big)$ & $\Big(c;d;c\Big)$ & $\Big(h,-g;e,f;g,h\Big)$ & $\Big(l,-k;i,j;k,l\Big)$ & $Amm2$ (\#38) \\
$\Big(a;b;c\Big)$ & $\Big(d;e;f\Big)$ & $\Big(g,h;i,j;k,l\Big)$ & $\Big(m,n;o,p;q,r\Big)$ & $Pm$ (\#6) \\
\hline
\end{tabular}
\caption{Candidate ground states for the free energy containing terms in all four order parameters $\mathbf{M}$, $\mathbf{L}$, $\mathbf{\Sigma}$, and $\mathbf{R}$. We use the notation $\mathbf{M}(a,b,c)$ for $m_1=a$, $m_2=b$, $m_3=c$; $\mathbf{L}(a,b,c)$ for $l_1=a$, $l_2=b$, $l_3=c$; $\mathbf{\Sigma}(a,b;c,d;e,f)$ for $\sigma_1=a+ib$, $\sigma_2=c+id$, $\sigma_3=e+if$; and $\mathbf{R}(a,b;c,d;e,f)$ for $r_1=a+ib$, $r_2=c+id$, $r_3=e+if$, consistent with the real order parameters $\mathbf{M} = (m_1,m_2,m_3)$ and $\mathbf{L} = (l_1,l_2,l_3)$, and the complex order parameters $\mathbf{\Sigma} = (\sigma_1,\sigma_2,\sigma_3)$ and $\mathbf{R} = (r_1,r_2,r_3)$, defined in the main text.}
\label{tab:subgroup_mlsr}
\end{table}

\begin{figure}
    \centering
    \includegraphics[width=0.99\linewidth]{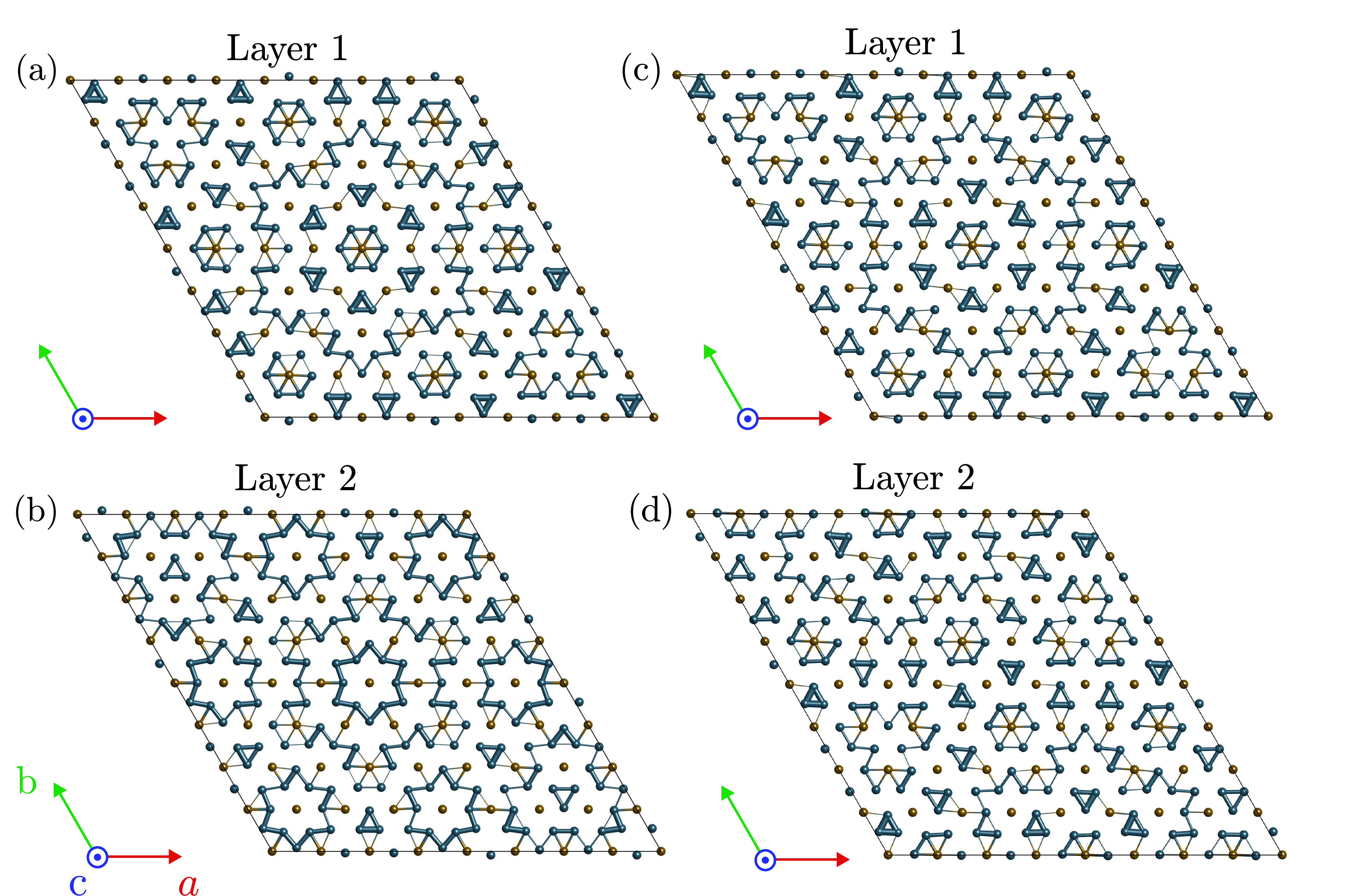}
    \caption{$8\times8\times2$ triple-$\mathbf{Q}$ phases of CsV$_3$Sb$_5$ obtained using the $\Sigma_1$ and $R_3$ phonon eigenvectors from DFPT. $(RRR)$ phase on the $z=0.25$ plane (a) and $z=0.75$ plane (b). $(\Sigma RR)$ phase on the $z=0.25$ plane (c) and $z=0.75$ plane (d).}
    \label{fig:RRR_SRR}
\end{figure}

In Table~\ref{tab:subgroup_r}, the most symmetric ground state originates from three components of the order parameter $\mathbf{R}$. The corresponding structure is shown in Fig.~\ref{fig:RRR_SRR}(a--b), and is referred to as the $(RRR)$ phase. Similarly, the $(\Sigma RR)$ phase is the analogue of the previously studied $(MLL)$ charge density wave~\cite{Christensen2021Dec,Ritz2023May} in the sense that the $\mathbf{\Sigma}$ order parameter spontaneously appears when the $\mathbf{R}$ order parameter is dominant because of the trilinear terms in Eq.~1 of the main text. Its structure is shown in Fig.~\ref{fig:RRR_SRR}(c--d). 

\clearpage